\documentclass[useAMS,usenatbib]{mn2e}
\usepackage{url}
\usepackage{amsmath}
\usepackage{times}
\usepackage{psfig}
\usepackage{graphics,graphicx}
\usepackage{lscape}
\usepackage{epsf}
\usepackage{epsfig}
\usepackage{color}
\usepackage{dcolumn}
\usepackage{natbib}
\bibpunct{(}{)}{;}{a}{}{,}
\def\aj{AJ}%
\def\araa{ARA\&A}%
\def\apj{ApJ}%
\def\apjl{ApJ}%
\def\apjs{ApJS}%
\def\aap{A\&A}%
\def\mnras{MNRAS}%
\def\pasp{PASP}%
\def\rmxaa{Rev. Mexicana Astron. Astrofis.}%
\def\nat{Nature}%
\def\bain{Bull.~Astron.~Inst.~Netherlands}%
\newcommand{\ka}{\mbox{WR\,102ka}}

\newcommand{\Lsun}{\mbox{$L_\odot$}}

\newcommand{\lsim}{\raisebox{-.4ex}{$\stackrel{<}{\scriptstyle \sim}$}}
\newcommand{\msim}{\raisebox{-.4ex}{$\stackrel{>}{\scriptstyle \sim}$}}
\newcommand{\mim}{\mbox{$\mu$m}}

\def\changed{}

\title[Isolated very massive star \ka]
{One of the most massive stars in the Galaxy
may have formed in isolation}
\author[Oskinova,  Steinke, Hamann, Sander, Todt, Liermann]
{L. M. Oskinova$^{1}$\thanks{E-mail: lida@astro.physik.uni-potsdam.de},  
M. Steinke$^{1}$, W.-R.
Hamann$^{1}$, A. Sander$^{1}$, H. Todt$^{1}$, A. Liermann$^{1,2}$\\
$^{1}$ Institute for Physics and Astronomy, University of Potsdam,
14476 Potsdam, Germany\\
$^{2}$ Leibniz-Institut f\"ur Astrophysik, Potsdam, An der Sternwarte 16, 14482 Potsdam, Germany}
\begin{document}

\date{Accepted . Received ; in original form 21.02.2011 22:32}

\pagerange{} \pubyear{2013}

\maketitle

\begin{abstract} Very massive stars, 100 times heavier than the sun, are
rare.  It is not yet known whether such stars can form in isolation or
only in  star clusters. The answer to this question is of fundamental
importance.  The central region of our Galaxy is ideal for investigating
very massive  stars and clusters located in the same  environment. 
{\changed We used archival infrared images to investigate the
surroundings of  apparently isolated massive stars presently known
in the Galactic Center. We find  that two such isolated  massive
stars display bow shocks and hence may be  ``runaways'' from their
birthplace. Thus, some isolated massive stars in the Galactic  Center
region might have been born in star clusters known in this region.
However,  no bow shock is detected around the isolated star \ka\ 
(Peony nebula star), which is one of the most  massive and luminous stars 
in the Galaxy. This star is located at the  center of an associated  
dusty circumstellar nebula. To study whether a star cluster may be ``hidden'' 
in the  surroundings of \ka, to obtain new and better spectra of this star, 
and to measure  its radial velocity, we obtained observations with the 
integral-field spectrograph  SINFONI at the ESO's Very Large Telescope 
(VLT). Our observations confirm that \ka\ is one of the  most massive stars 
in the Galaxy and reveal that this star is not associated with  a star cluster.  
We suggest that \ka\  has been born in relative isolation,  outside of
any massive star cluster.}     
\end{abstract}


\section{Introduction}

The stellar initial mass function (IMF) is the distribution of stellar 
masses in a population that formed together in one star-formation event 
on a spatial scale of up to a parsec \citep{kroupa2002}.   According to
the "random sampling" hypothesis,   massive stars may form occasionally
even in isolation, following a universal  probabilistic IMF
\citep{elm2006}. According to the alternative hypothesis  of "optimal
sampling", a very massive star can only form within a star  cluster
obeying a deterministic relation between the mass of the  cluster and
its most massive star \citep{Weidner2010,Weidner2013,kroupa2013}. 

Previous studies of massive stars located outside of star clusters 
considered only low-density environments, such as the Galactic field and
the  Magellanic Clouds, and presented support for an isolated star
formation  \citep{dewit2005,best2011,oey2013}. However, alternative
explanations were put forward arguing that each of the known isolated
stars was   formed in a cluster, but ejected from it during later
evolution \citep{gvar2012}.  Thus, the question about the origin of
isolated massive stars is not clarified yet. To  further progress in
understanding massive star formation, it is important to  investigate
different galactic environments. 

\begin{figure*}
\centering
\includegraphics[width=15cm]{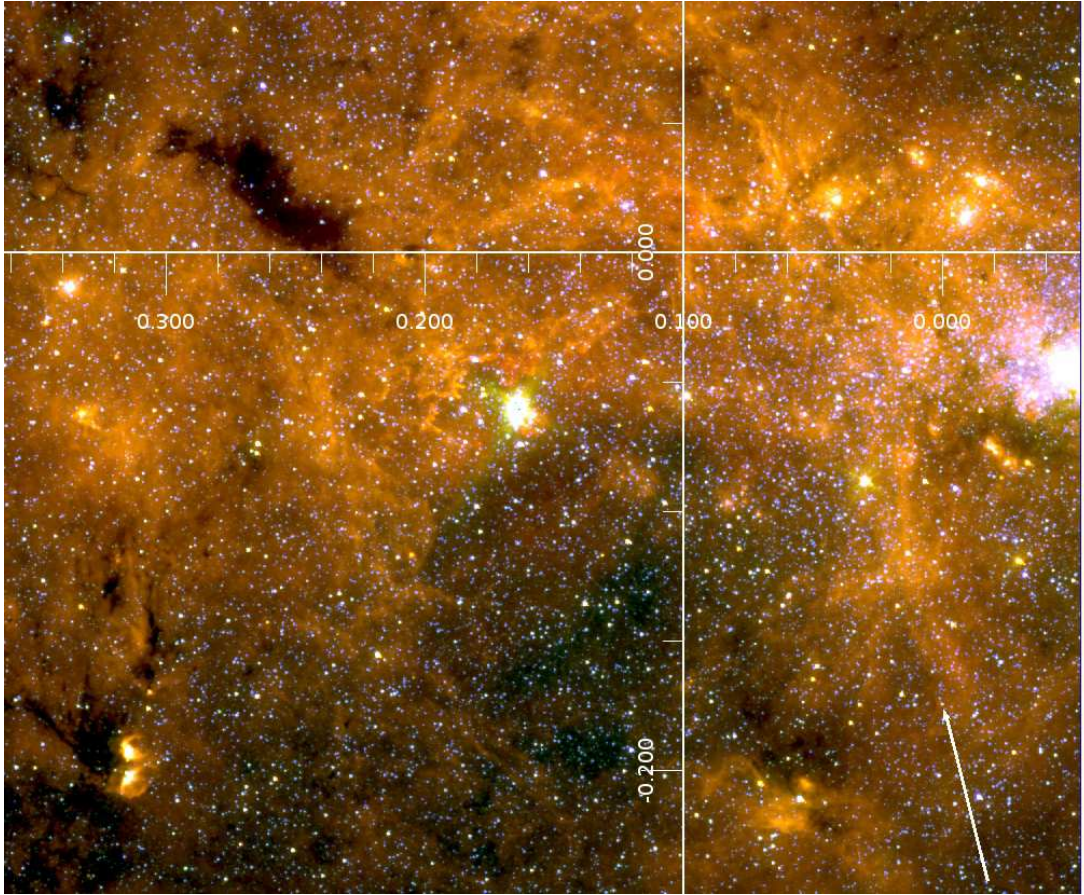}
\caption{Color composite {\em Spitzer} IRAC image of the GC: 
red  at $\sim 8$\,\mim, green at $\sim 5.8$\,\mim, blue at $\sim 3.6$\,\mim.
The image size is $24.8' \times 20.5'$  corresponding to 
$60\,{\rm pc} \times 50$\,pc at a distance of 8\,kpc. Galactic 
coordinates are indicated. 
The arrow at the lower right corner points to \ka. The 
length of the arrow is $250''$ ($\approx 10$\,pc). The bright cluster 
at the right edge of the image is the Central Cluster around Sgr\,A*.}
\label{fig:gcpeony}
\end{figure*}
%

The inner part of the Milky Way ($\approx 500$\,pc in Galactic
longitude and $\approx  60$\,pc in latitude) contains a large amount of
high-density molecular gas \citep{Ser1996}. This "Central Molecular
Zone" is characterized by large velocity dispersions, relatively high
temperatures, strong large-scale magnetic fields, and the proximity to
the super-massive black hole. Hence the star formation processes
occurring in this environment may be different from elsewhere in the
field of the Galaxy \citep{morris1996,Long2013}. This makes the
Galactic Center (GC) region an important test-bed for the theories  of
star- and cluster formation, where the universality of the IMF and 
massive star formation mechanisms can be investigated.  

Three very massive star clusters are located in the GC region.   The
Central Cluster envelopes the central super-massive black hole
coinciding with the radio source Sgr\,A*
\citep{Krabbe1995,Eckart1996,ghez1998}.  Two other massive star
clusters,  the Arches and the Quintuplet, are located  within 30\,pc
projected distance from Sgr\,A* \citep{Ser1998,Figer1999}.  While the
Arches cluster is younger and contains many OB and young Wolf-Rayet 
stars of WNL subtype  \citep{Mar2008}, the more evolved Quintuplet
cluster  (3-5\,Myr old) harbors many older Wolf-Rayet stars of
WC-subtype \citep{Tut2006,lho2012}.  Besides these compact stellar
conglomerates, many high-mass stars whose association  with  stellar
clusters is not obvious are scattered in the GC region
\citep{Cotera1999}.  The {\em Hubble Space Telescope} (HST) imaging  
survey of the inner $90\times 35$\,pc$^2$ of the Galaxy \citep{Wang2010} 
resulted in the detection of more than a hundred point-like sources of P$\alpha$ 
line excess. Large fraction of these sources was identified as massive stars 
located outside of the three known  stellar clusters \citep{Mauer2010, dong2011}. 
The origin of isolated massive stars in the GC is not known, but four scenarios may be envisaged. 
These stars 
({\it i}) were born in relative isolation \citep{Cotera1999}; 
({\it ii}) were formed within clusters that are dispersed by now; 
({\it iii}) were ejected from one of the three known clusters;
({\it iv}) are belonging to clusters which are not discovered yet. 

To investigate  the possible presence of a cluster around an apparently
isolated  massive star in this highly obscured region, deep infrared 
observations with  high-angular resolution are needed. From wide-field
infrared (IR) surveys alone,  it is impossible to find out whether these
massive stars are associated with clusters,  because of the high
stellar  density and the bright background in the GC region. 

For our study of apparently isolated massive stars in the Galactic 
Center region we select one of the most luminous and massive stars in the 
Galaxy, \ka\ (nicknamed the ''Peony nebula star'').  This star was 
discovered in 2002 \citep{Hom2003} at a projected distance of 19\,pc from 
the Central Cluster (see Fig.\,1). {\changed \citet{Bar2008} analysed the 
stellar spectrum obtained in 2002 combined with photometry. They  
revealed that \ka\ has an unconventionally high luminosity 
($\log{L}[\Lsun]=6.5\pm 0.2$) and concluded that its initial mass 
plausibly was in excess of $\sim 150\,M_\odot$. 
The goal of our new study, based on integral-field spectroscopy with
SINFONI, is twofold. Firstly, we analyze the K-band spectrum of \ka\ in
order to check if the new and better data confirm the exceptionally high
luminosity of this star. Secondly, we extract and classify all stellar
spectra from a mosaic of surrounding fields and check whether \ka\ is
accompanied by a yet unknown cluster of stars.}

The observations and data reduction is presented in section
\ref{sec:data}. The catalog and radial velocity measurements are
explained in section \ref{sec:cat}. The analysis of \ka's spectrum is 
presented in section \ref{sec:ka}, and the origin of this star is
discussed in section \ref{sec:orig}. The summary is given in section
\ref{sec:sum}.

\section{Observations and Data Reduction}
\label{sec:data}

\begin{figure}
\centering
\includegraphics[width=0.99\columnwidth]{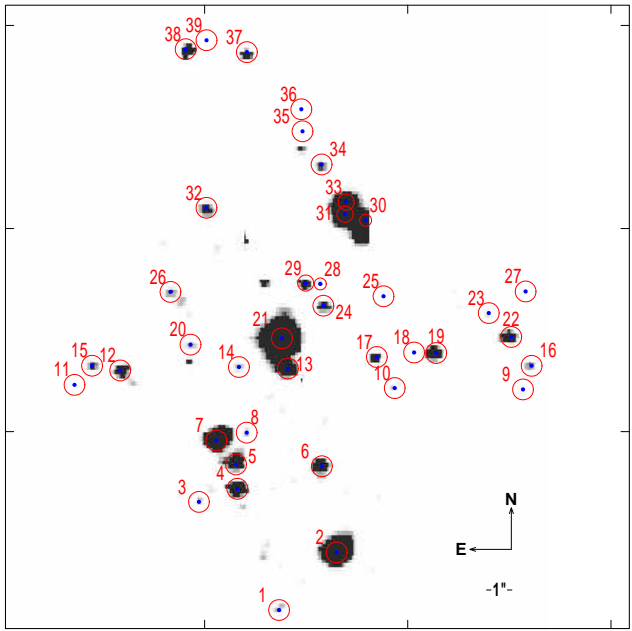}
\caption{Image of the collapsed grand mosaic cube. Open circles mark the 
detected sources with their running number in Table\,\ref{tab:starka}. 
The image size is $33.6''\times 38.5''$ ($1.3\,{\rm pc}\times 1.45\,{\rm pc}$). }
\label{fig:fov}
\end{figure}

The new data used in this work were obtained with the ESO VLT UT4
(Yepun) telescope in 2009 between May 15 and July 28.  The
observations were performed with the integral field spectrograph SPIFFI
of  the SINFONI module \citep{sinf2003}. {\changed This instrument delivers
a simultaneous, three-dimensional data-cube with two spatial dimensions
and one spectral dimension \citep{sinf2004}.}
The  K-band (1.95-2.45 $\mu$m) grating with resolving power R $\approx$
4000 was used. The spatial scale was chosen as   $0.25''$ per pixel,
giving a  field-of-view  of  $8\hbox{$^{\prime\prime}$ }\times
8\hbox{$^{\prime\prime}$ }$. The total observation consists of a mosaic of 11
pointings  (observational blocks or OBs) covering $\approx 33''\times
39''$ ($\approx 1.3$\,pc $\times$ 1.5\,pc)  centered on \ka\ 
(see Fig.\,\ref{fig:fov}). The finally reduced data consist of 3-D 
data cubes with one spectral and two spatial dimensions.

The adaptive optics facility could not be used because there is no 
sufficiently bright reference  star in the neighborhood.
The log of observations is given in Table\,\ref{tab:log}. 
ABBA   (science field -- sky -- sky
-- science field) cycles were performed at each pointing.  To obtain 
flux calibrated spectra,  standard stars were observed at similar 
airmass and in the same mode as the science targets.  During our
observations the seeing was between $0.6''$ and $1.1''$.
{\changed Thus, the angular resolution of our observations is limited by 
seeing.}

%

The raw data  were cleaned with the tool L.A.Cosmic \citep{vanDokkum2001}
and  a self-written tool for removing artifacts and hot pixels. The
further  reduction of the data was primarily performed with the ESO
pipeline tool ESORex,  version 3.9.6 \citep{Modigliani2007}. 


\begin{table*}
\caption{Log of observations}
\label{tab:log}
\begin{center}
\begin{tabular}{lcccccrr}
\hline
\hline
\noalign{\vspace{1mm}}
 OB   & date & R.A. (J2000) & DEC (J2000)            & exp.     & sky
& \multicolumn{1}{c}{standards}  & average\\
  (ID)   &  2009 & 17$^{\rm h}$46$^{\rm m}$ & $-29^{\circ}01'$ & time [s]   &
$^{(1)}$  &HIP$^{(2)}$ & seeing [$''$] \\
\noalign{\vspace{1mm}}
\hline

361412 & May 15 & $18\fs 23$ & $21\farcs 1$& 150 & B1 & 88857
& 0.63 \\
 361414 & May 15 & $19\fs 04$ &  $28\farcs 3$& 150 & B2 & 88857
& 1.04 \\
 361416 & May 28 & $18\fs 48$ &  $28\farcs 3$ & 150 & B1 & 85223
& 1.10 \\
 361418 & June 29 & $17\fs 94$ &  $28\farcs 7$ & 30 & B2 & 88857
& 1.07 \\
 361420 & June 27 & $18\fs 76$ &  $37\farcs 1$& 150 & B1 &
88857 & 1.02 \\
 361422 & June 27 & $18\fs 21$ &  $37\farcs 0$ & 10 & B1 & 88201
& 0.91 \\
 361424 & June 27 & $17\fs 67$ &  $37\farcs 1$ & 150 & B1 &
88857 & 0.78 \\
 361426 & June 27 & $17\fs 12$ &  $37\farcs 2$ & 150 & B2 &
88857 & 0.91 \\
 361428 & June 29 & $18\fs 39$ &  $44\farcs 2$& 150 & B1 &
88201 & 0.97 \\
 361430 & June 29 & $17\fs 85$ &  $44\farcs 2$ & 150 & B2 &
105164 & 0.90 \\
 361432 & June 29 & $17^{\rm s}91$ &  $51\farcs 4$ & 150 & B1 &
88857 & 1.04 \\
\hline
\end{tabular}
\end{center}
\small{
$^1$: Sky fields B1 and B2  were centered at the coordinates
$17^{\rm h}46^{\rm m}7\fs 22$, $-29^{\circ}02'54\farcs 4$ and 
$17^{\rm h}46^{\rm m}29\fs 47{\rm s}$, $-28^{\circ}59'14\farcs 2$,
respectively. \\
$^2$: {\em Hipparcos} catalog names  of standard stars}
\end{table*}


Sky subtraction for the science observations was performed using 
the standard pipeline The wavelength
calibration was done by using a Ne-Ar lamp. 

Each science observation was flux-calibrated.  Early BV-type stars were
observed as calibration sources,  because  their spectra  are relatively
featureless in the K-band.   The  model fluxes of these B-type stars
were taken from PoWR models (see section \ref{sec:ka}). Accidentally,  a
foreground B-type star (number 38 in Table\,\ref{tab:starka}) 
is present  in our science field.  The 2MASS  photometry of this star was used
to fine-tune our final flux calibration.

The flux-calibrated  3-D data cubes of the individual fields were 
combined to a grand mosaic cube. This cube was ``collapsed'', i.e.\ summed
over  all wavelengths for the purpose of point source detection. The   
point spread function (PSF) was obtained from our observations of
standard stars and applied for the source detection in the science
fields. We securely detect 39 point sources in the observed field (see
Table\,\ref{tab:starka} and Fig.\,\ref{fig:fov}).

Our observations are sensitive down to an apparent magnitude of $K_{\rm
s}=14.8$\,mag. With the reddening towards \ka\  \citep{Bar2008}, the
extinction in the K-band amounts to about 3\,mag.  Thus, at the distance
of 8\,kpc, our observations are sensitive to absolute K-band magnitudes
$<-2.8$\,mag. 

{\changed We considered the possibility that stars fainter than \ka\ in the
K-band and located within $\sim 1''$ from \ka\ may remain undetected. Especially, 
hotter objects would not be easily distinguishable from imaging and photometry
alone. Fortunately, we can use the combination of imaging, photometry and line
spectrum.
We estimate that a contribution of 10\%\ from OB stars to the K-band spectrum
of \ka\ would not escape from our detection because of their He- and Br$\gamma$ 
line features. How many OB stars can be hidden below this limit?
The sensitivity of our observations (K=14.8\,mag) is 6\,mag ($\sim 250$ times) 
fainter than \ka\ (K=8.7, see Table\,\ref{tab:starka}) and corresponds to a
main-sequence star of about 15 $M_\odot$ \citep[based on the evolutionary 
tracks by][]{Brott2011}. Thus, about 25 such B-type stars could be hidden
within the point-spread function of \ka. However, more massive stars 
would be easier to detect. Consider, for instance, an O star of 
40\,$M_\odot$ and $T_{\rm eff}=30$\,kK. Such star would have an absolute 
K magnitude of $-5.1$\,mag, which is 3.7\,mag or a factor of 20 
fainter than \ka. Thus, a maximum of two such $40\,M_\odot$ stars could 
be outshined by \ka. }

\begin{table*}
  \small
  \begin{centering}
\caption{Catalog of the detected stellar sources}
  \vspace{2mm}
  \begin{tabular}{lrrrrrrr}
\hline\hline
No. & RA        & DEC      & $K_{\rm s}^\ast$ &Spectral& Lumin.&$V_{\rm rad}$
\rule[0mm]{0mm}{4mm}\\ 
    & 17$^{\rm h}$46$^{\rm m}$  & $-29^{\circ}01'$& [mag]         &type    &
class & [km/s]      \\
\hline
1  & $18\fs 0981$ & $54\farcs 8277$  & 14.0   & K3	& II	  & 105  \\
2  & $17\fs 8638$ & $51\farcs 2778$  & 10.4   & M5-6	& I-II    & 30   \\
3  & $18\fs 4277$ & $48\farcs 1673$  & 14.2   & K5	& II	  & 15   \\
4  & $18\fs 2739$ & $47\farcs 3639$  & 12.3   & M3	& I-II    & -25  \\
5  & $18\fs 2739$ & $45\farcs 8601$  & 12.8   & M2	& I	  & 130  \\
6  & $17\fs 9224$ & $45\farcs 9700$  & 12.7   & M1	& I	  & -100 \\
7  & $18\fs 3545$ & $44\farcs 3839$  & 11.2   & M4	& I	  & 225  \\
8  & $18\fs 2300$ & $43\farcs 9032$  & 14.0   & K1	& I-II    & 40   \\
9  & $17\fs 1021$ & $41\farcs 2459$  & 14.5   & K2	& II	  & 275  \\
10 & $17\fs 6221$ & $41\farcs 1566$  & 14.3   & M0	& II-III  & 15   \\
11 & $18\fs 9478$ & $41\farcs 0399$  & 15.8   & late$^\dagger$    \\
12 & $18\fs 7500$ & $40\farcs 0786$  & 13.3   & K5	& I-II    & 35   \\
13 & $18\fs 0615$ & $39\farcs 9550$  & 12.2   & M1	& I	  & -10  \\
14 & $18\fs 2666$ & $39\farcs 8657$  & 14.1   & K0	& I-II    & 100  \\
15 & $18\fs 8672$ & $39\farcs 7833$  & 13.8   & M0	& II	  & 260  \\
16 & $17\fs 0654$ & $39\farcs 7902$  & 13.9   & M1	& II	  & 50   \\
17 & $17\fs 6953$ & $39\farcs 2477$  & 13.5   & M2	& II	  & -100 \\
18 & $17\fs 5488$ & $38\farcs 9731$  & 14.8   & K2	& II-III  & 135  \\
19 & $17\fs 4536$ & $39\farcs 0005$  & 12.5   & K3	& I	  & 55   \\
20 & $18\fs 4644$ & $38\farcs 4856$  & 14.0   & K3	& I-II    & 135  \\
21 & $18\fs 0908$ & $38\farcs 0942$  &  8.7   & WN9-10  &	  & 60   \\
22 & $17\fs 1460$ & $38\farcs 0118$  & 12.8   & M2	& I	  & -75  \\
23 & $17\fs 2412$ & $36\farcs 5424$  & 14.3   & K2	& II	  & 40   \\
24 & $17\fs 9150$ & $36\farcs 0892$  & 13.1   & M1	& I-II    & 65   \\
25 & $17\fs 6733$ & $35\farcs 5055$  & 14.6   & K3	& II	  & 170  \\
26 & $18\fs 5449$ & $35\farcs 2377$  & 13.8   & K0	& I	  & 20   \\
27 & $17\fs 0874$ & $35\farcs 2583$  & 14.6   & late$^\dagger$  \\
28 & $17\fs 9297$ & $34\farcs 7502$  & 14.2   & K4	& II	  & -5   \\
29 & $17\fs 9883$ & $34\farcs 6953$  & 13.2   & M0	& I-II    & -5   \\
30 & $17\fs 7466$ & $30\farcs 8295$  & 11.3   & M4	& I	  & 130  \\
31 & $17\fs 8271$ & $30\farcs 4587$  & 12.1   & M1	& I	  & 130  \\
32 & $18\fs 3984$ & $30\farcs 0604$  & 13.4   & M2	& II	  & 0	\\
33 & $17\fs 8271$ & $29\farcs 7102$  & 11.3   & M4	& I-II    & 130  \\
34 & $17\fs 9224$ & $27\farcs 3962$  & 13.6   & K4	& I-II    & 160  \\
35 & $18\fs 0029$ & $25\farcs 3500$  & 14.9   & K4	& II-III  & 65   \\
36 & $18\fs 0103$ & $23\farcs 9973$  & 14.8   & K2	& III	  & -105 \\
37 & $18\fs 2300$ & $20\farcs 4749$  & 12.9   & K3	& I	  & -30  \\
38 & $18\fs 4790$ & $20\farcs 2963$  & 12.0   & B3-5	& IV-V    & 0	\\
39 & $18\fs 3984$ & $19\farcs 7470$  & 14.8   & M0	& III	  & 150  \\
\hline
\end{tabular}
  \end{centering}
\\
\rule{0mm}{4mm}
{\small $^\ast$ Magnitudes derived from our 
flux-calibrated spectra, using the 2MASS filter transmission 
function \citep{2mass2006}}\\
$^\dagger$ Too faint for closer classification
\label{tab:starka}
  \end{table*}

\section{Catalog, spectral classification, and radial velocities measurements}
\label{sec:cat}

\begin{figure}
\centering
\includegraphics[width=\columnwidth]{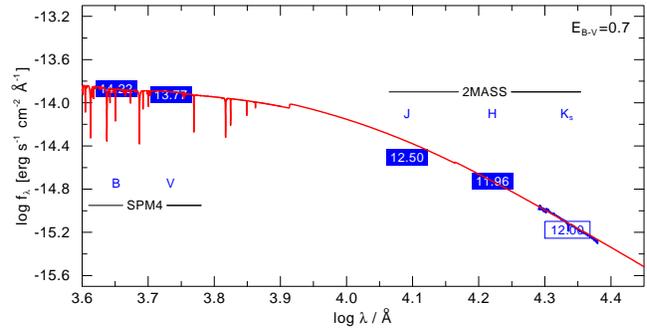}
\caption{ 
Spectral energy distribution of star No.\,38 (alias 
SPMID\,5080099266). The blue boxes indicate the B, V  
and 2MASS photometric magnitudes. The SINFONI K-band spectrum 
can be seen as a blue thin line.
Over-plotted as thick red line is the synthetic spectrum 
from the PoWR model with $T_{\rm eff}=16$\,kK and 
$\log{L/L_\odot}=3.2$ for a distance of 2.5\,kpc 
}
\label{fig:bspec}
\end{figure}
%

The flux-calibrated spectra  of all stars were extracted from the data
cubes,  and their spectral types were determined.   With two exceptions,
all stars have stellar spectra of late types.  A closer classification
was obtained from the CO absorption bands, specifically  $^{12}{\rm CO}$
and  $^{13}{\rm CO}$, adopting  standard criteria
\citep{goor1994,lho2009}. 

We detect only two early-type stars in the field: \ka\ and 
a star of spectral type B3-5 (No.\,38 in Table\,\ref{tab:starka}). 
The catalog search revealed  the latter star is an optical source
(SPMID\,5080099266) of  V=13.77\,mag \citep{Girard2011}.  We
complemented its K-band spectrum  obtained with SINFONI with available
photometric measurements. These  data were fitted with a PoWR model
spectrum,  and stellar parameters were, thus, obtained (Fig.\,\ref{fig:bspec}).  
The star has much lower reddening than  the GC region, and is therefore
undoubtedly a foreground object at a distance of about  $\sim 2.5$\,kpc.
Its  effective temperature is $T_{\rm eff}\approx 16$\,kK.   
\begin{figure}
\centering
\includegraphics[width=0.99\columnwidth]{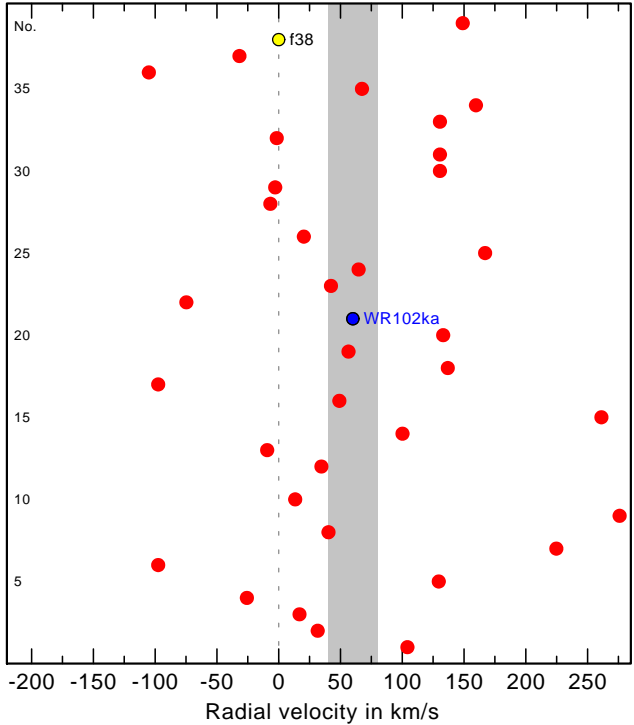}
\caption{Radial velocities of the stars in our SINFONI field. 
The vertical axis gives the source number from 
Table\,\ref{tab:starka}. The shaded area is centered on 
$V_{\rm rad}({\rm WR\,102ka}) = 60$\,km\,s$^{-1}$, while its 
width roughly corresponds to the uncertainty of the measurements.}
\label{fig:radvel}
\end{figure}
%

Our observations would also be capable  to detect massive young stellar
objects (MYSO) if they were present. However,  none of the stars  in our
sample shows spectral features characteristic for MYSOs, such as
emissions from CO bands or Br$\gamma$  \citep{Bik2006}. 

The radial velocities of \ka\ and the other stars  were determined from
the  Doppler-shift of prominent stellar lines in their spectra. We
estimate that the Doppler shifts are accurate to about
40\,km\,s$^{-1}$,  which corresponds to slightly more than one 
detector pixel in wavelength.

Radial velocities of late-type stars are determined from the first three
band heads of the $^{12}$CO  $\lambda 2.293$\,\mim\ (2-0),  $\lambda
2.322$\,\mim\ (3-1) and $\lambda 2.352$\,\mim\ (4-2)  molecular
transitions \citep{Gorlova2006, GF2008},  where e.g.\ (2-0) stands for the change
in the vibrational quantum number of the molecule from $\nu=2$ to
$\nu=0$. These three measurements were averaged, and the results are
given in Table\,\ref{tab:starka} and Fig.\,\ref{fig:radvel}. 
The range of radial velocities  of the
late-type stars detected in our observed field is similar to that in 
other nearby areas of the GC region \citep{lho2009}. 

\begin{figure*}
\centering
\includegraphics[width=1.3\columnwidth]{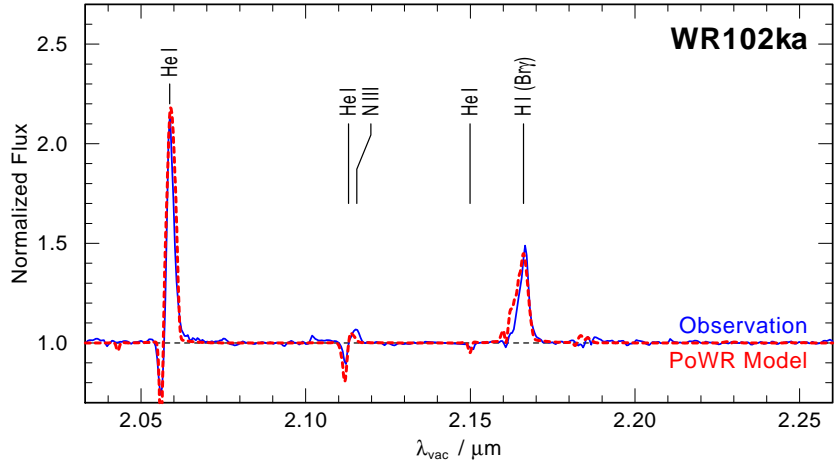}
\caption{Observed SINFONI spectrum of \ka\ (blue solid line) together with 
best-fitting PoWR model spectrum (red dashed line) with
luminosity $\log{L_{\rm bol}[L_\odot]}=6.47\pm0.05$, hydrogen abundance 
$X({\rm H})=30 \pm 5$\%, mass-loss rate  
$\log{\dot{M}[M_\odot\,{\rm yr}^{-1}]}=-4.6\pm 0.1$, 
terminal wind speed $v_\infty=400$\,km\,s$^{-1}$ and
radial velocity  60\,km\,s$^{-1}$.}
\label{fig:specka}
\end{figure*}

Stellar wind lines are not symmetric. To determine  the radial
velocity of \ka, we therefore compared our model to the observation (see
Figure\,\ref{fig:specka} and Section \ref{sec:ka}). 
{\changed We applied different radial velocity shifts to the model spectrum
and calculated the difference to the observation. The $\chi^2$ sum has a
pronounced minimum for a radial velocity of 60\,km\,s$^{-1}$. The visual
inspection confirms that this value provides a consistent fit for all
prominent lines in the spectrum.

Profile shapes might depend on details of the stellar-wind model. As a test
we calculated models with different wind velocity laws. The wind model with
$v_\infty = 400$\,km\,s$^{-1}$ and the standard $\beta$-velocity exponent
$\beta=1$ reproduces the line shapes best. Thus the wind velocity parameters
do not infer a large uncertainty to the radial velocity determination. 

We also compared our observed  spectrum with that obtained in 2002 
by \citet{Hom2003}. No change in  radial velocity is detectable between 
these two observations separated by about seven years. }

The radial velocity of the B-type star (No.\,38) is obtained from the
hydrogen Br$\gamma$ line, and is found to be
0\,km\,s$^{-1}$. This is in line with our finding that this star is 
in the foreground.

There is also diffuse  Br$\gamma$ emission all over our observed field.  We
selected star-free areas  and extracted the nebular spectra. The
wavelength  of the nebular Br$\gamma$  emission is un-shifted,  i.e.\ the
nebular gas has the same radial velocity as the B-type star  No.\,38. We
conclude that this diffuse emission comes from an H\,{\sc ii} region
that is also located in the foreground, and possibly related to the
foreground B-star. According to our model, this star produces
$\log{\Phi}[{\rm s}^{-1}]=43.5$ hydrogen ionizing photons.
Unfortunately, this foreground H\,{\sc ii} region  contaminates our
observations of the circumstellar nebula around \ka. 

\begin{figure}
\centering
\includegraphics[width=0.8\columnwidth]{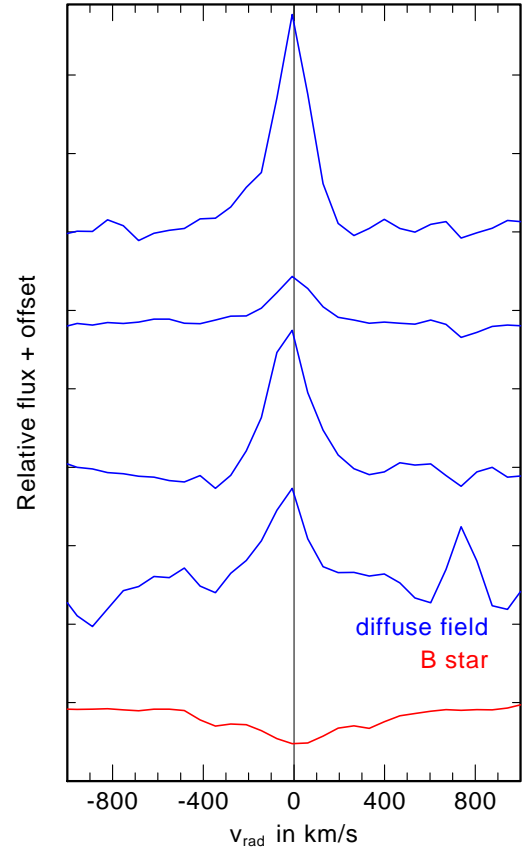}
\caption{Observed Br$\gamma$ line in the spectrum of the B-type
star No.\,38  (red line), compared to the nebular lines at different
star-free locations in our field (blue lines), displaced by
arbitrary offsets. }
\label{fig:nebline}
\end{figure}
%

Our analysis shows that there are no indications for a group of 
stars co-moving  with  \ka\ which would have  been expected for 
a star cluster \citep[e.g.][]{lho2009}. Thus, we conclude that 
our observations do not detect a star cluster associated with \ka.

\section{Spectral analysis of \ka}
\label{sec:ka}

{\changed The earlier spectral analysis of \ka\ \citep{Bar2008} was
based  on a spectrum obtained in 2002. We found that the star has an
unconventionally high luminosity ($\log{L}[\Lsun]=6.5\pm 0.2$) at a
relatively low stellar temperature ($T_\ast=25.1$\,kK). The spectral
type of \ka\ was determined as WN10. \citet{Bar2008} showed that \ka\
is located above the Humphreys-Davidson limit in the
Hertzsprung-Russell  Diagram (HRD) \citep[see Figure\,16
in][]{Bar2008}. Such stars may be  unstable and show large variability
changing their location in the HRD on the time-scale of years. In
order to investigate whether the stellar parameters of \ka\ remained
stable over more than 7\,yr, to check for radial velocity changes (as
may be expected in a binary star), and to verify the outstanding
stellar parameters derived in the previous work, we obtained and
analyzed the new spectrum.
}

\begin{figure}
\centering
\includegraphics[width=0.7\columnwidth,angle=-90]{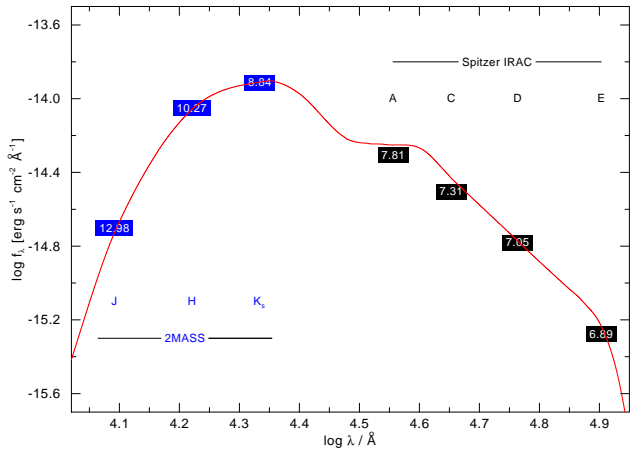}
\caption{Spectral energy distribution of WR 102ka. The boxes correspond
to the photometric magnitudes (as printed in the boxes) measured by
2MASS and  {\em Spitzer} IRAC. The red line is the continuum from a 
PoWR model with  $T_{\rm eff}=25$\,kK and $\log{L/L_\odot}=6.47\pm0.15$ 
at distance modulus and extinction of $D\!M = 14.6$\,mag and $E_{\rm
B-V}=8.1$\,mag, respectively.}
\label{fig:sed}
\end{figure}
 
As in the previous work, the observed spectrum of \ka\  was analyzed by
comparison with PoWR model atmospheres \citep{Ham2004}. The PoWR code
has been used extensively to analyze the  spectra of massive stars in
the IR as well in the ultra-violet and optical range
\citep{osk2007,lho2010, osk2011,Sander2012}. The PoWR code solves the
non-LTE radiative transfer in a spherically expanding atmosphere
simultaneously with the statistical equilibrium equations and accounts
at  the same time for energy conservation. Complex model atoms with
hundreds  of levels and thousands of transitions are taken into
account. The computations for the present paper include detailed model
atoms for hydrogen, helium, carbon, oxygen, nitrogen, and silicon. Iron
and iron-group elements with millions of lines are included through the
concept of super-levels \citep{Graf2002}. The extensive inclusion of
the iron group elements is important because of their blanketing effect
on the  atmospheric structure.

\begin{figure}
\centering
\includegraphics[width=\columnwidth]{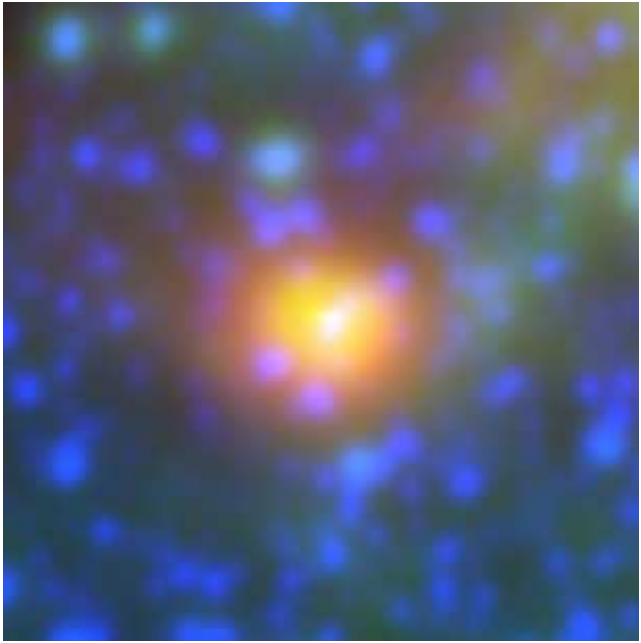}
\caption{Color composite WISE image of \ka.
Red is at $\sim 22$\,\mim, green is at $\sim 12$\,\mim, blue is 
at $\sim 3.4$\,\mim.  Image size is 
$3.6' \times 3.6'$. North is up and east is to the left.}
\label{fig:wise}
\end{figure}

Each stellar atmosphere model is defined by its effective temperature,
surface gravity, luminosity, mass-loss rate, wind velocity, and chemical
composition. The gravity determines the density structure of the
stellar atmosphere below and close to the sonic point. From the
pressure-broadened profiles of photospheric lines, the spectroscopic
analysis allows to derive the gravity and thus the stellar mass.

{\changed The analysis consists of two coupled steps, the fit of the
normalized
line spectrum (Fig.\,\ref{fig:specka}) and the fit of the spectral
energy distribution (Fig.\,\ref{fig:sed}). The line fit confirms the
parameters obtained in the previous analysis by \citet{Bar2008}: effective
temperature $T_\ast = 25.1$\,kK (referring to the radius of Rosseland
optical depth 20), hydrogen mass fraction $X_{\rm H}=30 \pm 5$\%, terminal
wind velocity $v_\infty=400$\,km\,s$^{-1}$, mass-loss rate
$\log{\dot{M}[M_\odot\,{\rm yr}^{-1}]}=-4.7\pm 0.1$ (for a clumping
parameter of $D=6$).

When comparing the spectral energy distribution of the model with
photometric observations, we tested various extinction curves that are
available for  the near IR
\citep{Cardelli1989,Moneti2001,nish2009}.  Fortunately, the parameters
obtained by the fit ($\log L$, $E_{\rm B-V}$) do not differ
significantly. The model shown in  Fig.\,\ref{fig:sed} is reddened with
the Moneti  et\,al.\ extinction curve.   Finally, we obtain $\log
L/L_\odot = 6.47 \pm 0.15$,  thus confirming the very high stellar
luminosity of \ka.

It is known that very luminous massive stars (such as luminous blue
variables) display bolometric luminosity variations
\citep[e.g.][]{Clark2009,Shol2011}.  The 2MASS measurements employed in
Fig.\,\ref{fig:sed} were obtained on 2000/10/07.  Our calibrated
SINFONI spectrum of \ka\ obtained in 2009 (see Table\,\ref{tab:log})
gives a slightly higher flux than this 2MASS photometry. The difference
corresponds to 0.1\,mag and probably just reflects the
uncertainty of the calibrations. Nevertheless, it is  interesting to
note that \ka\ was reported to vary in the H- and J-band  by 0.16\,mag
\citep{Matsu2009}.

\citet{Bar2008} speculated that two marginal spectral features which
are merely visible in their data might be attributed to N\,{\sc v} and
He\,{\sc ii}, and thus possible
originate from a hotter un-resolved companion. The same weak features
seem also to be present in our new spectrum, but the identification
remains questionable.

The new data thus verify that the Peony star is one of the most
luminous and massive stars known in the Galaxy. According to the latest
set of very massive star evolutionary tracks \citep{Yusof2013}, its
initial mass was about $150\,M_\odot$, while its current mass amounts
to about $100\,M_\odot$ and its age is $\sim 2$\,Myr. Using the
mass-luminosity relation for very massive stars from \citet{graf2011},
the current mass of \ka\ is $\approx110\,M_\odot$. }

\section{On the origin of \ka.}
\label{sec:orig}

\begin{figure*}
\centering
\includegraphics[width=1.9\columnwidth]{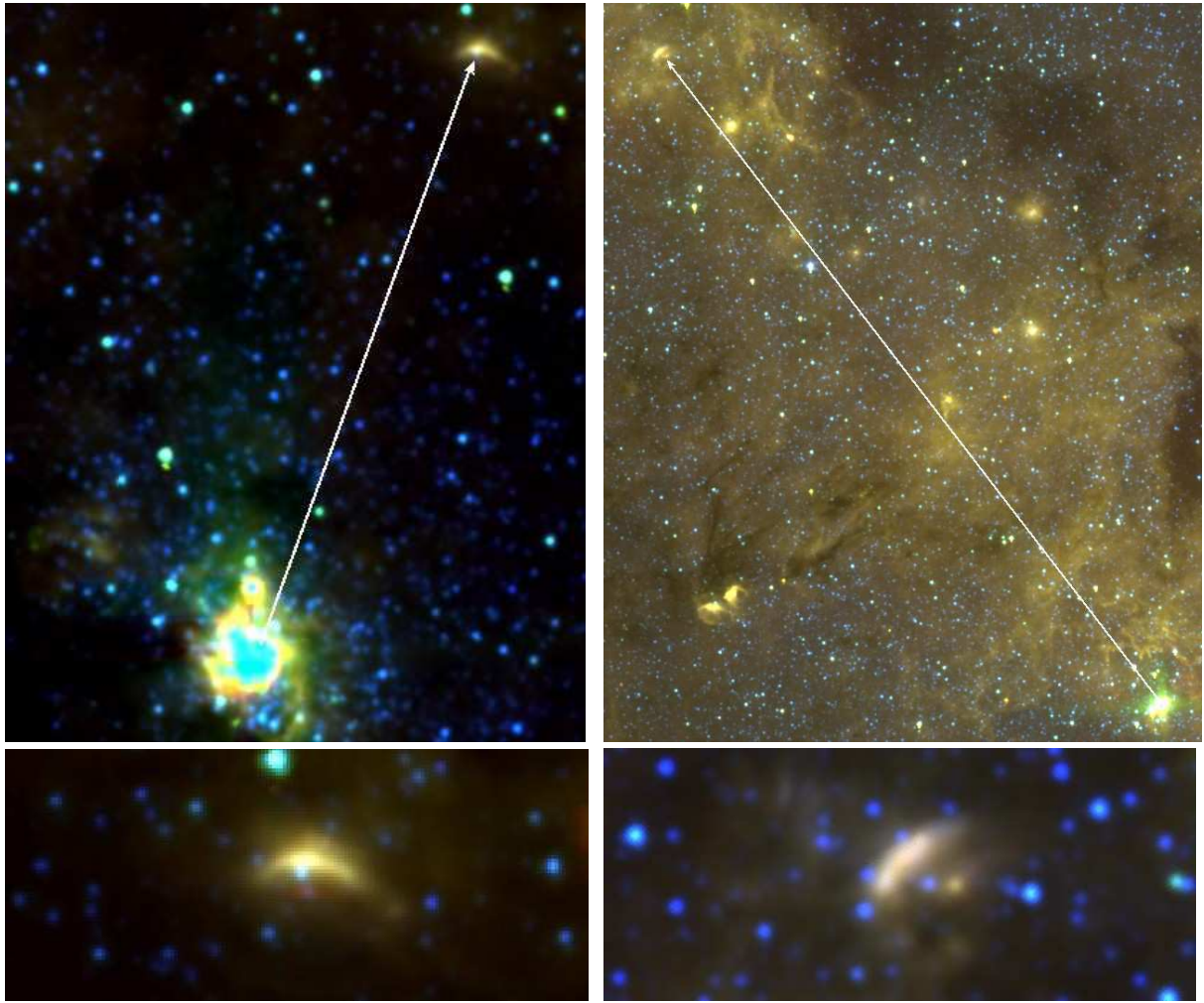}
\caption{Two bow-shocks seen in the color composite {\em Spitzer} 
IRAC images of the GC region. Red is at $\sim 8$\,\mim, green is  
at $\sim 5.8$\,\mim, and blue is  at $\sim 3.6$\,\mim.
North is up, and east is to the left.
{\it Left panels:} The arrow (length $4.4'$) points 
to the bow shock around the \mbox{O4-6}-type supergiant CXOGC\,J174532.7-285617.
The bright overexposed object at the bottom is the Central Cluster 
around Sgr\,A*. The lower panel zooms on the bow shock. 
{\it Right panels:} The arrow 
points the bow shock around the WN7-8h star CXOGC\,J174712.2-283121. Its
length ($22.2'$) indicates its separation from the  Quintuplet cluster. 
The lower panel zooms on the bow shock.}
\label{fig:bow1}
\end{figure*}
%

The hypothesis of optimal IMF sampling implies that a star as massive as 
\ka\  was  born in a massive star cluster \citep{Weidner2013}. To compare our 
observations with this  hypothesis we used the {\em Mcluster} code, 
which is a publicly available tool for initializing star cluster models  
and binary-rich stellar populations \citep{kup2011}. The simulations predict 
that $\approx$\,300 stars with masses exceeding $20\,M_\odot$ shall be present
in a cluster initially containing a star with the mass of \ka. Among 
these stars $\approx 10$ may  be more massive than $100\,M_\odot$.  These
predictions are in strong disagreement with our observational results. 

{\changed If \ka\ would reside in a star cluster with a mass predicted 
by the simulations, such cluster would be clearly detected in our 
observations. The characteristic size of a very massive young star cluster 
is $\sim 1$\,pc, while angular 
resolution  of our observations is $\msim 0.024$\,pc (at the distance of 
the GC). As can be seen in Fig.\,\ref{fig:fov}, \ka\ is the brightest 
object in the observational mosaic image. On the detector it covers 
approximately $10\times 10$\,pixels. This corresponds to an area with 
$\approx 0.1$\,pc in diameter. Assuming that a hypothetical cluster 
is ultra-compact with a diameter of 0.1\,pc its stellar density would have been 
$>10^6\,M_\odot$\,pc$^{-1}$. This is an order of magnitude above stellar 
densities in the most massive clusters in the Local Group 
\citep[e.g.][]{PZ2010} and is unrealistic. Also, the spectrum of \ka\ is
well described as a originating in a wind of a WR star, with no indications 
that  it is a composite spectrum from many OB and WR stars. 
We are therefore certain that no massive star cluster is associated with \ka.}

We now investigate the possibility that \ka, albeit  isolated at present, 
was formed within a star cluster $\sim 2$\,Myr ago. Two scenarios shall 
be considered: its parental star cluster has dissolved, 
or \ka\ was ejected from one of the known (or not yet known) massive star
clusters. 

The first scenario is relatively easy to dismiss. The self-consistent 
dynamic N-body star-cluster model which accounts for the important tidal 
field of the Galaxy predicts that the massive clusters in the GC region  
are not yet dissolved at the age of \ka\ \citep{PZ2002}. An example 
of an older massive star cluster is the Quintuplet \citep{lho2012}.

The ejection scenario may seem more likely. Even a massive star can leave 
its parental star cluster if it gains sufficiently high  spatial velocity 
either as a result of a supernova explosion in a binary system \citep{Bla1961},
or by close stellar encounters during the early dynamical evolution 
of the cluster  \citep{Amb1954, Allen1974, gies1986}.  Such stars are 
commonly termed ``runaways''. 

{\changed The radial velocity of \ka\ is $+60\pm 20$\,km\,s$^{-1}$. 
This is similar to the radial velocities of the atomic and molecular gas 
at these galactocentric distances that is found to be in the interval
between $\msim 10$\,km\,s$^{-1}$ and $\lsim +100$\,km\,s$^{-1}$  
\citep{Martin2004,An2013}. However, the radial velocity of \ka\ 
is lower than the mean radial velocity of the stars in the Quintuplet
Cluster,  $+113$\,km\,s$^{-1}$ \citep{lho2009}, or the Arches
Cluster,  +95\,km\,s$^{-1}$ \citep{figer2002}. }

Mid-IR 24\,\mim\ images  of the circumstellar nebula around \ka\ were 
obtained with the {\em Spitzer} 
telescope\footnote{\url{
http://www.nasa.gov/mission_pages/spitzer/news/spitzer-20080715.html}}. 
Figure\,\ref{fig:wise} shows the WISE image of this circumstellar dusty 
nebula (``Peony nebula'') heated by the stellar radiation of its 
central star \ka. It was suggested that the nebula contains stellar material 
that was lost  by \ka\ during previous evolutionary stages 
\citep{Clark2005,Bar2008}.
The central position  of the star in its nebula shows that the star
basically remained at the same location during its recent evolution.

Taking the radial velocity as a lower limit to the
spatial velocity, \ka\ traveled at least 130\,pc during its life-time.
This is much more than the distance between \ka\ and the  massive star
clusters in the GC. Is one of these clusters a possible birthplace
for \ka? The Quintuplet and the Central cluster are older than the age
of \ka\ assuming single star evolution (2\,Myr), and therefore it is
not likely that the star was formed there. The Arches cluster is
younger and might be sufficiently massive for the formation of \ka.
However, it seems not likely that just the most massive star has been stripped
off from the Arches cluster while many less massive stars remained bound. 
Summarizing, we conclude that \ka\ was not formed in any of
these clusters.

{\changed Stars with  supersonic velocities relative to the ambient 
matter  tend to form a bow shock in the direction of its motion 
\citep{vanBuren1995,Comeron1998,Moffat1998}. These bow shocks can be 
detected with IR imaging  \citep[e.g.][]{kob2010,Peri2012}.  We inspected 
the IR images of other isolated massive stars in the  GC region.
Here we report the apparent presence of bow shocks around two 
isolated  massive stars in the Galactic center region,
CXOGC\,J174532.7-285617 (O4-6I) \citep[alias P\,114,][]{dong2011} and 
CXOGC\,J174712.2-283121 (WN7-8h), see Figure\,\ref{fig:bow1}. 
Previously and independently, these bow shocks were noticed in \citet{Mauer2010}, 
who reported shell-like or bow-shock like structures around a handful of 
massive stars in the GC. If the nature of these IR-bright bow shocks is 
confirmed by future  observations, it would indicate that the cluster ejection 
mechanism operates in the GC region. 
 
On the other hand, the IR {\em Spitzer} IRAC image of \ka\  
(Fig.\,\ref{fig:pirac}), no obvious circumstellar structure 
resembling a bow shock is seen. This presents an additional argument
supporting the suggestion that \ka\ resides at a place of its original
formation or not far from it. 

\begin{figure}
\centering
\includegraphics[width=\columnwidth]{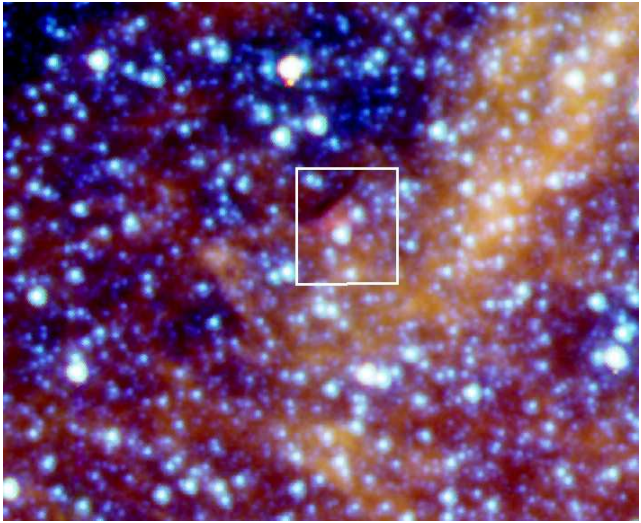}
\caption{Color composite {\em Spitzer} IRAC image of \ka.
Red is at $\sim 8$\,\mim, green is at $\sim 5.8$\,\mim, blue is 
at $\sim 3.6$\,\mim. The white box illustrates 
the area covered in our SINFONI observation. Image size is 
$2.9' \times 3.6'$. North is up and east is to the left.}
\label{fig:pirac}
\end{figure}
The majority of isolated massive stars in the GC do not display obvious bow
shocks either. On the basis of observational evidence we have today,  we
shall conclude} that the massive star population in the GC region
consists of a heterogeneous mixture of stars in clusters, stars which
were ejected or stripped from clusters, and stars which were formed
outside of clusters.  Overall, our results show that at least one very
massive star, \ka, is not associated with a star cluster, contrary to
the prediction of the  IMF optimal sampling hypothesis. The GC region
apparently provides an  environment where massive stars can form in
clusters as well as in relative  isolation.   

\section{Summary}
\label{sec:sum}
We obtained the integral field spectroscopic observations of the very 
luminous and massive WR-type star \ka\ located in the GC. {\changed 
We confirm previous conclusions that this object is a very massive star 
with initial mass $\msim 150\,M_\odot$ and current mass $\sim 100\,M_\odot$.} 
On the basis of\\ 
\smallskip
 ({\it i}) the absence of early type stars within $\msim 1$\,pc from \ka, \\
 ({\it ii}) the absence of a group of stars of similar age and comoving with \ka, \\
 ({\it iii}) the radial velocity of \ka\ which is similar to the typical
galactocentric velocities at this location, \\
 ({\it iv}) the presence of dusty  circumstellar nebula around \ka, which was
ejected in previous evolutionary stage of \ka, but containing \ka\
in its center, \\
 ({\it v}) the absence of a bow shock around \ka, while
bow shocks are apparently present around some other massive stars in the GC,\\
\smallskip
we conclude that one of the most massive and luminous stars in the
Galaxy, \ka\ (Peony Nebula star) may have formed in relative isolation. 

\section*{Acknowledgments}   
Based on observations collected at the ESO VLT (program 383.D-0323(A)).
This work has used observations obtained with the Spitzer Space 
Telescope, which is operated by the Jet Propulsion Laboratory,
California Institute of Technology under a contract with NASA. 
This work has extensively used the NASA/IPAC Infrared Science Archive, 
the NASA's Astrophysics Data System, and the SIMBAD database, operated at 
CDS, Strasbourg, France. This publication makes use of data products 
from the Two Micron All Sky Survey, which is a joint project of the University 
of Massachusetts and the Infrared Processing and Analysis Center/California 
Institute of Technology, funded by the National Aeronautics and Space
Administration and the National Science Foundation. This publication makes 
use of data products from the Wide-field Infrared Survey Explorer, which 
is a joint project of the University of California, Los Angeles, and the 
Jet Propulsion Laboratory/California Institute of Technology, funded by 
the National Aeronautics and Space Administration. 
{\changed We are grateful to the referee for useful comments which helped 
to improve this paper}. 
Funding for this research 
has been provided by DLR grant 50\,OR\,1302 (LMO) and DFG grant HA\,1455/22 (AS). 


\end{document}